\documentstyle[sprocl,amssymb,epsfig]{article}

\def\kms{~km\,s$^{-1}$}

\newcommand{\mnras}[1]{MNRAS}
\newcommand{\apj}[1]{ApJ}
\newcommand{\apjs}[1]{ApJS}
\newcommand{\apjl}[1]{ApJL}
\newcommand{\nat}[1]{Nature}
\newcommand{\aap}[1]{A\&A}
\newcommand{\araa}[1]{ARA\&A}
\newcommand{\aaps}[1]{A\&ASS}
\newcommand{\aj}[1]{AJ}
\newcommand{\apss}[1]{Ap\&SS}

\begin{document}

\title{\Large \bf White Dwarfs: Contributors and Tracers of the Galactic 
	Dark--Matter Halo}

\author{L.V.E. Koopmans \& R.D. Blandford}

\address{Caltech, mailcode 130--33, Pasadena CA 91101, USA}


\maketitle

\abstracts{We examine the claim by Oppenheimer et al. (2001) that the
local halo density of white dwarfs is an order of magnitude higher
than previously thought. As it stands, the observational data support
the presence of a kinematically distinct population of halo white
dwarfs at the $>$99\% confidence level. A maximum-likelihood analysis
gives a radial velocity dispersion\footnotemark\ of $\sigma^{\rm
h}_{U}$=150$^{+80}_{-40}$\kms and an asymmetric drift of $v_{\rm
a}^{\rm h}$=176$^{+102}_{-80}$\kms, for a Schwarzschild velocity
distribution function with $\sigma_U$:$\sigma_V$:$\sigma_W$=1:2/3:1/2.
Halo white dwarfs have a local number density of
$1.1^{+2.1}_{-0.7}\times 10^{-4}$~pc$^{-3}$, which amounts to
0.8$^{+1.6}_{-0.5}$ per cent of the nominal local dark-matter halo density
and is 5.0$^{+9.5}_{-3.2}$ times (90\% C.L.) higher and thus only marginally 
in agreement with previous estimates. We discuss
several direct consequences of this white-dwarf population
(e.g. microlensing) and postulate a potential mechanism to eject young
white dwarfs from the disc to the halo, through the orbital
instabilities in triple or multiple stellar systems.}

\footnotetext{Hereafter, if not otherwise indicated, 
errors give the 90\% statistical confidence level.} 

\section{Introduction}

Recently, Oppenheimer et al. (2001; O01 hereafter) found 38 white
dwarfs (WD) in a sample of 99 high proper motion WDs, which were
claimed to have kinematics inconsistent with that of both the old
stellar disc {\sl and} the thick disc, and therefore form a newly
discovered halo population with an inferred density at least an order
of magnitude higher than previously thought (e.g. Gould et al. 1998).
Reid et al. (2001) have criticized this result and conclude that these
WDs form the high-velocity tail of the thick disc, based on a
comparison of their $U$--$V$ velocity distribution with that of
M-dwarf stars. Unfortunately, neither conclusion is statistically
supported.

In this proceeding, we discuss a maximum-likelihood analysis of the
complete sample of 99 WDs and derive the local phase-space density of
thick-disc {\sl and} halo WDs. This then allows us to draw more robust
conclusions.  For a discussion of the selection of the sample, we
refer to O01. Details of the likelihood analysis and phase-space
density estimate can be found in Koopmans \& Blandford (2001; KB01
hereafter). For clarity, throughout the proceeding the low- and
high-velocity dispersion components are referred to as the thick-disc
and halo population, respectively. This does not imply that all WDs
could not have originated from the thin or thick disc (e.g. Hansen
2001; KB01).

\section{The White--Dwarf Velocity Distribution Function}

We model the local ($\vec{x}_0$) velocity distribution function (VDF)
as a superposition of two Schwarzschild (i.e.~Maxwellian) VDFs,
i.e.~$f_{\rm B}(\vec{x}_0,\vec{v})$. We assume a constant space 
density of WDs throughout the surveyed volume. Each component has two free
parameters, the radial velocity dispersion ($\sigma_U$) and the
asymmetric drift ($v_{\rm a}$). In addition, the parameter $r_n$ is
the ratio of the thick-disc to halo WD number densities ($n$). We
assume that the vertex deviations are zero and that the ratios of the
radial ($U$), azimuthal ($V$) and vertical ($W$) velocity dispersions
of the ellipsoidal VDFs are
$\sigma_U$:$\sigma_V$:$\sigma_W$=1:2/3:1/2, in agreement with
observations (see KB01 for references).  The probability that an
observed WD with a velocity vector on the sky, $\vec{p}=(v_l,v_b)$, is
drawn from this VDF becomes $P = \int_\infty^\infty (v_l^2 +
v_b^2)^{3/2} f_{\rm B}(\vec{v}=\vec{p}+ v_r \hat{r})\, {\rm d}v_r/
\int_{\cal V} (v_l^2 + v_b^2)^{3/2} f_{\rm B}(\vec{v}=\vec{p}+ v_r
\hat{r}) {\rm d}^3 v$, where $v_r$ is the radial velocity along the
line-of-sight and $v_l$ and $v_b$ are the velocities projected on the
sky in Galactic coordinates. This assumes that the WDs are all proper-motion 
limited ($\sim$90\% are). ${\cal V}$ is then the velocity space in
which the WDs could have been found, given the restriction that
${v_l^2 + v_b^2>(\mu_0 r)^2}$, where r is the distance to the WD and
$\mu_0$ is the survey's lower limit on the proper motion
($\mu_0\sim$0.33\,$''$\,yr$^{-1}$). For the $\sim$10\% magnitude limited
WDs, we find that their maximum detection volumes ($V_{\rm max}$) are nearly
identical. For these, we modify the likelihood function, although the 
differences in parameter estimates are 
within the errors quoted below. By varying the five free parameters and optimizing
the log-likelihood, ${\cal L}$=$\sum_i \log(P_i)$, of the sample, we
solve for their most likely values and their error range. The results
(excluding 3 thin-disc WDs) are: (a) $\sigma_{U,\rm
d}$=62$^{+8}_{-10}$\kms, (b) $\sigma_{U,\rm h}$=150$^{+80}_{-40}$\kms,
(c) $v_{a,\rm d}$=50$^{+10}_{-11}$\kms, (d) $v_{a,\rm
h}$=176$^{+102}_{-80}$\kms and (e) $r_n$=16$^{+30}_{-11}$ (KB01).

\section{The Local Halo White--Dwarf Density}

To normalize the local phase-space density of halo plus thick-disc
WDs, i.e. $n^{\rm td+h}_{\rm 0,WD}\cdot f_{\rm B}(\vec{x}_0,\vec{v})$,
it remains to estimate their local density, $n^{\rm td+h}_{\rm 0,WD}$.
We do this in the conventional way, by summing the $1/V_{\rm max}$
values of the halo plus thick-disc WDs, where $V_{\rm max}$ is the
smallest of the two volumes in which a WD could have been detected,
when limited either by its proper motion or by its magnitude
(e.g. O01; KB01). We find that $\sim$90\% of the WDs are proper-motion
limited and three WDs are very likely associated with the thin
disc. The latter WDs are removed from the sample, as we are only
interested in the halo plus thick-disc density, for which we
determined the VDF. We find $n^{\rm td+h}_{\rm 0,WD}=(1.9\pm0.5)\times
10^{-3}$~pc$^{-3}$, of which $n^{\rm h}_{\rm 0,WD}=
1.1^{+2.1}_{-0.7}\times 10^{-4}$~pc$^{-3}$ belongs to the halo,
given $r_n$ found from the maximum-likelihood analysis of the
sample. The local halo plus thick-disc density estimate agrees within
$\sim$1~$\sigma$ with the estimate from Reid et al. (2001).  
The halo density is 5.0$^{+9.5}_{-3.2}$ times higher than the best
previous estimate of ${\tilde n}^{\rm h}_{\rm 0,WD}=2.2\times
10^{-5}$~pc$^{-3}$ from Gould et al.~(1998), inferred
from the mass function of halo stars. Locally, our density estimate
amounts to 0.8$^{+1.6}_{-0.5}$\% of the nominal local dark-matter
halo density of $8\times 10^{-3}$~M$_\odot$\,pc$^{-3}$ (Gates et
al. 1995). 

\section{The Global Halo White--Dwarf Density}

Given the {\sl local} phase-space density of halo plus thick-disc
WDs and a potential model for the Galaxy, we can estimate the {\sl
global} white-dwarf density in a large part of the Galactic
halo, using Jean's theorem (e.g. May \& Binney 1986). For simplicity,
we assume a spherical logarithmic potential, $\Phi(r)=v_{\rm c}^2
\ln(r/r_{\rm c})$, with rotation velocity $v_{\rm c}$=220\kms\ and a
solar radius of $r_{\rm c}$=8~kpc. For the halo WDs, which are less
affected by the disc/bulge potentials this is probably a
reasonable assumption, although this does not hold for the
lower-velocity thick-disc WDs. A analytic expression for the
phase-space density of halo WDs can then be derived for any point in
halo that is in dynamic contact with the Solar neighborhood (KB01) .
From this phase-space density model, we find an oblate
($q=(c/a)_\rho\sim0.9$) distribution of halo WDs with a total mass
inside 50~kpc of $\sim$2.6$\times$$10^9$~M$_\odot$ and a radial mass
profile $n(r)\propto r^{-3.0}$. We expect $q$ to decrease further if
a proper flattened Galactic potential is used. This halo WD mass
amounts to $\sim$0.4\% of the total Galactic mass inside 50~kpc,
$\sim$4\% of the Galactic stellar mass, or 
$\Omega_{\rm WD}\sim 10^{-4}$.
Including disc WDs, the total mass of WDs in
the Galaxy is $\sim$9\% of the stellar mass, in agreement with
standard population synthesis models (Hansen 2001; KB01).

\section{The LMC Microlensing Optical Depth}

Given our analytic expression of the halo WD phase-space density, we
can integrate over the line-of-sight towards the Large Magellanic
Cloud (LMC) to estimate the microlensing optical depth of this
population. We find $\tau_{\rm WD}^{\rm h}\sim 1.3\times 10^{-9}$,
which is $\sim10^2$ times lower than observed (Alcock et
al. 2000). The thick-disc WDs have a $\sim$3 times higher optical
depth.  The integrated halo plus thick-disc WD optical depth is
therefore still 1--2 orders of magnitude too small.

\section{Conclusions, Discussion \& Future}

Modeling of the local phase-space density of halo plus
thick-disc WDs indicates that there is a population of
pressure-supported halo WDs with a {\sl local} density
5.0$^{+9.5}_{-3.2}$ times higher than previously estimated,
although globally ($r\lesssim50$~kpc) the mass contribution is
negligible ($\sim$0.4\%). Both the low- and high-velocity populations
have similar color, magnitude and ages distributions (Hansen 2001; KB01).
Two particularly interesting surveys to search
for more high proper motion WDs are: (i) {\sl a wide survey with the
Advanced Camera for Surveys (ACS)} on the Hubble Space Telescope,
which we estimate should find $\sim$5 WDs (KB01) and (ii) a similar
survey as that by O01 towards the Galactic anti-center to constrain
the WD velocities perpendicular to the Galactic plane.

The white-dwarf data set from O01 and results derived from it
(i.e. our most likely model) have met a number of interal and external
consistency checks (KB01) and agree with {\sl all} observational
constraints (e.g. direct observations, metal pollution of the ISM,
microlensing, etc.) known to us.  The color-color relation of the WDs
shows them to be quite young (Hansen 2001) and have a birth rate
roughly that expected from Galactic population synthesis models
(KB01). Why then do a large fraction of these WDs have such high
spatial velocities?  {\sl To explain this we postulate the following;
the majority of high-velocity WDs in the Galactic halo originate from
multiple (i.e. $N\ge 3$) stellar systems in the Galactic disc. In case
of a striple stellar system, stellar evolution (i.e. mass loss and/or
transfer) of the (probably more massive) inner binary stars changes
the ratios of orbital periods and eventually destabilizes the
system. At that point, the lightest star, a WD in the inner binary, is
ejected from the system into the halo -- possibly through a slingshot
of the outer star -- leaving behind a recoiled compact binary (KB01).}
This postulate implies a direct relation between the radial surface
density of the stellar disc, the star-formation rate and population
synthesis models, which all seem to be in agreement (KB01), {\sl but}
still need considerable study.

\section*{Acknowledgments}
We thank Ben Oppenheimer for valuable discussions and sending their
tabulated results. We thank David Graff and 
Andy Gould for pointing out an error in the normalisation of our 
likelihood function. LVEK thanks Priya Natarajan for organising a
productive and well-organised meeting.
This research has been supported by NSF~AST--9900866 and 
STScI~GO--06543.03--95A.

\end{document}